# Low-coherence optical diffraction tomography using a ferroelectric liquid crystal spatial light modulator

CHANSUK PARK,[1,2] KYEOREH LEE,[1,2] YOONSEOK BAEK,[1,2]
AND YONGKEUN PARK,[1,2,3,*]

[1]*Department of Physics, Korea Advanced Institute of Science and Technology (KAIST), Daejeon 34141, South Korea*
[2]*KAIST Institute for Health Science and Technology (KIHST), KAIST, Daejeon 34141, South Korea*
[3]*Tomocube, Inc., Daejeon 34051, South Korea*
[*]*Corresponding author: yk.park@kaist.ac.kr*

**Abstract:** Optical diffraction tomography (ODT) is a three-dimensional (3D) label-free imaging technique. The 3D refractive index distribution of a sample can be reconstructed from multiple two-dimensional optical field images via ODT. Herein, we introduce a temporally low-coherence ODT technique using a ferroelectric liquid crystal spatial light modulator (FLC SLM). The fast binary-phase modulation provided by the FLC SLM ensures a high spatiotemporal resolution with considerably reduced coherent noise. We demonstrate the performance of the proposed system using various samples, including colloidal microspheres and live epithelial cells.

## 1. Introduction

Optical diffraction tomography (ODT) is a three-dimensional (3D) label-free quantitative phase imaging technique [1, 2]. The 3D refractive index (RI) distribution of a sample can be reconstructed from multiple two-dimensional (2D) optical field images via ODT. To this end, an inverse scattering problem based on either the first-order Born or Rytov approximations has to be solved. Because ODT enables the observation of intact biological samples in a noninvasive and quantitative manner, it has been actively utilized in various fields, including cell biology [3-5], immunology [6], pharmacology [7], biotechnology [8], and nanotechnology [9].

In ODT, the 3D RI distribution is reconstructed from multiple 2D optical field images containing both the amplitude and phase information of the sample [6]. To measure optical field images, interferometric microscopes with coherent illumination sources are generally utilized. Multiple 2D optical field images can be obtained by rotating the sample [10, 11] or scanning the sample axially [12-15], or by angle scanning of an illumination beam [16-18].

A major technical challenge in ODT is coherent noise, including parasitic fringe and speckle pattern. In general, highly coherent light sources are used for illumination in ODT. Although a coherent source is required for the straightforward retrieval of phase information via interferometry, it inevitably produces coherent noise because of its long coherence length. The coherent noise results from multiple reflections or scattering from various optical components and dust particles. The coherent noise significantly decreases the quality of the reconstructed tomograms in ODT.

To address coherent noise, various approaches have been employed. Subtracting the background can reduce the coherent noise as well as aberration [19-21]. However, owing to the time-varying noise caused by various mechanical instabilities or the source itself, coherent noise cannot be completely eliminated. Recently, data-driven approaches have been employed to reduce coherent noise [22, 23]; however, they require training with a considerable amount of data.

The introduction of a spatiotemporally low-coherence source is a fundamental solution for eliminating coherent noise [24-26]. However, the implementation of a spatiotemporally low-coherence source is a challenging task. This is because the interference condition over a large field of view is difficult to maintain, particularly when the illumination angle of the beam is varied [27, 28].

In ODT, galvanometric mirrors [16, 29] and digital micromirror devices (DMDs) [18, 30] are commonly used to scan the angle of the illumination beam impinging onto a sample. However, for low-coherence light galvanometric mirrors cannot be used because of the decoherence from mirror-based angle tilting [28]. Such decoherence during illumination can be avoided by adopting a diffraction geometry using a DMD. However, the use of a DMD induces spectral dispersion due to its inherent echelle grating structure. The spectral dispersion can be compensated by introducing a secondary DMD as a dispersion compensation unit [28]; however, a highly complicated alignment procedure should be implemented when two DMDs are employed. Moreover, the energy efficiency becomes extremely low as the light experiences lossy DMD diffraction twice before it reaches the sample. A nematic liquid crystal on silicon spatial light modulator is a suitable option in terms of the diffraction efficiency [31]. However, the relatively low framerate restricts its application to the fast dynamics of live cells.

Herein, we present a method for fast- and low-noise ODT employing a temporally low-coherence light source and a diffraction-based illumination scanning method using a ferroelectric liquid crystal spatial light modulator (FLC SLM). The FLC SLM provides binary-phase modulation, which is utilized for generating temporally multiplexed sinusoidal patterns. The sample is illuminated with this patterned beam, from which diffracted components at three illumination angles are digitally retrieved. Using the method, we achieved high-quality reconstructions of 3D RI tomograms with high speed and energy efficiency. To demonstrate its efficacy, we measured the 3D RI distribution of epithelial cells and time-lapse 3D images of chimeric antigen receptor (CAR) T and K562 cancer cells.

## 2. Method

A schematic of the optical setup is presented in Fig. 1(a). A superluminescent light emitting diode (SLED) (EXS210099-03, Exalos, Switzerland) was used as an economic temporally low-coherence and spatially coherent light source. The center wavelength and bandwidth of the SLED were 450 nm and 6 nm, respectively. The corresponding coherence length was 37.5 µm. The short coherence length helps avoid interference between multiple reflected beams from optical components such as cover slips. This interference is a major source of coherent noise [28].

The collimated beam from the SLED was split into a sample and a reference arm using a polarizing beam splitter (PBS1). The illumination unit, which modulated the scanning angle, consisted of a PBS2, half-wave plate (HWP), and FLC SLM (M150, 4.5 kHz framerate, Fourth Dimension Display, U.K.). The FLC SLM operated in a binary-phase mode. The two states of a single pixel on the FLC SLM had two different orientations of the FLC optic axis denoted as $\mathbf{e_1}$ and $\mathbf{e_2}$ [Fig. 1(b)]. The FLC layer served as a quarter-wave plate (QWP), which is equivalent to a HWP for a round trip. The polarization of the input beam $\mathbf{E^{in}}$ was defined by PBS2 ($\mathbf{p^{in}}$). The HWP aligned $\mathbf{p^{in}}$ and $\mathbf{p^{out}}$ such that they bisected the angle between two states, $\mathbf{e_1}$ and $\mathbf{e_2}$ [32]. For each binary state, the FLC layer rotated the reflected polarization ($\mathbf{E_1}$ and $\mathbf{E_2}$) by an angle $\pm 2\theta$, where $\theta$ is the angle between $\mathbf{E^{in}}$ and both $\mathbf{e_1}$ and $\mathbf{e_2}$. The output polarizations $\mathbf{E_1^{out}}$ and $\mathbf{E_2^{out}}$ were the projections of $\mathbf{E_1}$ and $\mathbf{E_2}$ to the readout polarization $\mathbf{p^{out}}$. The output beam had a binary phase modulated between 0 and π, i.e., $\mathbf{E_1^{out}} = -\mathbf{E_2^{out}}$. The beam reflected from the FLC SLM was then scattered by the sample and collected by the objective lens. High numerical aperture (NA) water immersion lenses (UPLSAPO 60XW, NA of 1.2, Olympus, Japan) were used for both the illumination and collection parts of the setup.

To maintain coherence between the beams of both arms, a dispersion compensation unit was placed on the reference arm [the dashed box in Fig. 1(a)]. The unit consisted of a QWP, PBS3, glass block, and metallic mirror mounted on a translation stage. A translation stage was used to tune the optical path length difference between the sample and the reference arm. A 90-mm-thick glass block (E-BK7) was placed in the reference arm to compensate for the remaining high-order dispersion. The beams from both arms were combined by PBS4 and formed a spatially modulated interferogram on an imaging sensor. The use of a FLC SLM enables a more compact and simpler setup than the DMD-based illumination unit used in Ref. [28].

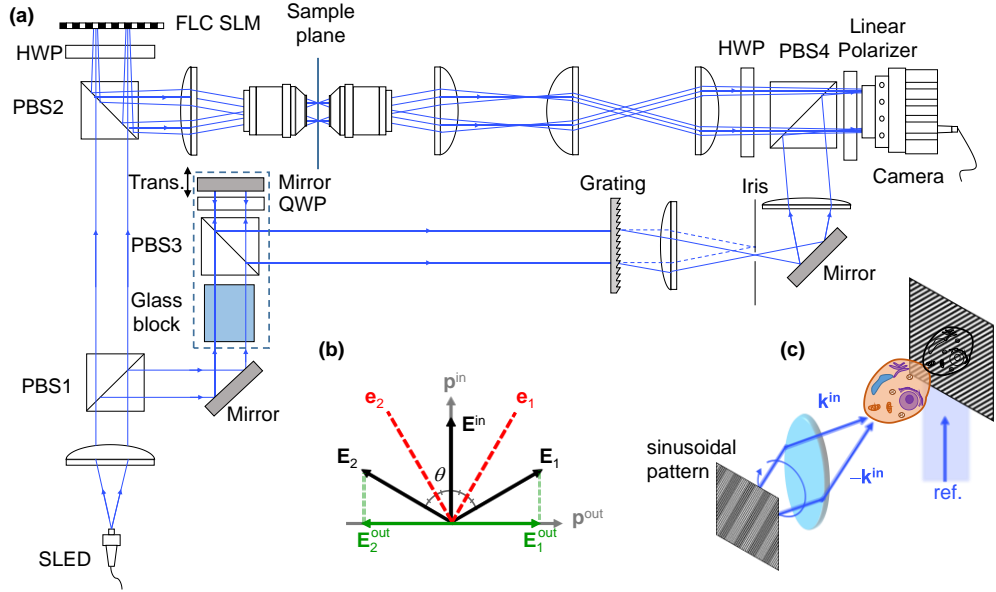

Fig. 1. (a) Schematic of the setup. PBS, polarization beam splitter; HWP, half-wave plate; QWP, quarter-wave plate; trans, translation stage. (b) Principle of binary-phase modulation using FLC SLM. For each state, the optic axis of the FLC layer is tilted by $\pm\theta$ ($\mathbf{e_1}$ and $\mathbf{e_2}$). The tilted FLC layer serves as a HWP to the input beam $\mathbf{E^{in}}$ and rotates it by an angle of $\pm 2\theta$. ($\mathbf{E_1}$ and $\mathbf{E_2}$). After reflection by the PBS, the output beams are $\mathbf{E_1^{out}}$ and $\mathbf{E_2^{out}}$. (c) Illustration of the ODT angle-scanning procedure: A sinusoidal pattern with various $\mathbf{k^{in}}$ is illuminated on the sample, and the resulting interferogram from the sample-scattered beam and the reference beam recorded by the camera.

A reference beam was tilted by a diffraction grating (NT46-067, Edmund, U.S.). A telescopic 4-*f* system behind the grating further magnified the diffracted reference beam to match the sampling condition for off-axis holography. An iris selectively allowed the first-order diffraction from the grating to pass through. A high-speed camera (CS1, 3 kHz framerate, PCO, Germany) was used to maximize the imaging speed.

Illumination at a systemically controlled oblique angle was realized by uploading a calculated sinusoidal pattern in the form of cos($M\mathbf{k}^{in}\cdot\mathbf{x}$) to the FLC SLM [Fig. 1(c)], where $\mathbf{k}^{in}$, $\mathbf{x}$, and $M$ represent the transverse component of the oblique illuminating wave vector, spatial coordinate on the FLC SLM plane, and magnification factor ($M = 1/133$) from the FLC SLM plane to the sample plane, respectively. Because the FLC SLM resulted in binary-phase modulation, a modified digital time-multiplexing method was employed [30]. A single sinusoidal pattern was uploaded on the FLC SLM as four decomposed binary (1 and −1) patterns. Scattered fields from each binary pattern were recorded and numerically summed with a corresponding weight factor [Fig. 2(a)].

In principle, illumination by a sinusoidal pattern is equivalent to simultaneous illumination by two plane waves exp($i\mathbf{k}^{in}\cdot\mathbf{x}$) and exp($-i\mathbf{k}^{in}\cdot\mathbf{x}$). However, in practice we found that a residual normal incidence ($\mathbf{k}^{in} = 0$) remained because of incomplete FLC SLM modulation. We adopted a structured illumination scheme [33] to decouple the three complex fields. Each sinusoidal pattern cos($M\cdot\mathbf{k}^{in}\cdot\mathbf{x} + \phi_n$) was displayed three times with different additional phases $\phi_1 = 0$, $\phi_2 = 2\pi/3$ and $\phi_3 = 4\pi/3$. The scattered field from a sinusoidal pattern with a phase delay $\phi_n$ is given by $E_n = E_+\exp(i\phi_n) + E_-\exp(-i\phi_n) + E_0$ [Fig. 2(b)], where $E_\pm$ and $E_0$ refer to the scattered fields from exp($\pm i\mathbf{k}^{in}\cdot\mathbf{x}$) and a normal incidence term, respectively. By measuring the three fields $E_1$, $E_2$, and $E_3$, we can obtain the plane wave illumination scattered fields ($E_+$ and $E_-$) for each $\mathbf{k}^{in}$. Fifty sinusoidal patterns with different $\mathbf{k}^{in}$ for oblique illumination and a single normal pattern were used to obtain 101 = 50 × 2 + 1 complex fields. Equivalently, a total of 601 (= 50 × 4 × 3 + 1) interferograms were recorded.

To reconstruct the 3D RI tomograms from the multiple measured interferograms, we applied the ODT algorithm, followed by the field retrieval algorithm [2, 34]. The ODT algorithm is based on the Fourier diffraction theorem:

$$\tilde{U}(\mathbf{k}) = \frac{i}{2k_z}\tilde{S}(\mathbf{k} - \mathbf{k}^{in}) \qquad (1)$$

where $\tilde{U}(\mathbf{k})$ and $\tilde{S}(\mathbf{k})$ are the Fourier spectra of the retrieved complex field and the sample scattering potential, respectively; further, $k_z$ is the axial component of the illumination beam wave vector. By mapping the complex field images obtained with various illumination angles, the scattering potential of the sample $S(\mathbf{r}) = k_0^2[n^2(\mathbf{r}) - n_m^2]$ was reconstructed [2]. Here, $k_0$ is the wave vector in vacuum, $n(\mathbf{r})$ is the 3D RI distribution of the sample, and $n_m$ is the RI of the surrounding medium. The detailed ODT algorithm and its MatLab implementation, regularization algorithm to address the missing cone issue, and field retrieval algorithm are described elsewhere [34-36].

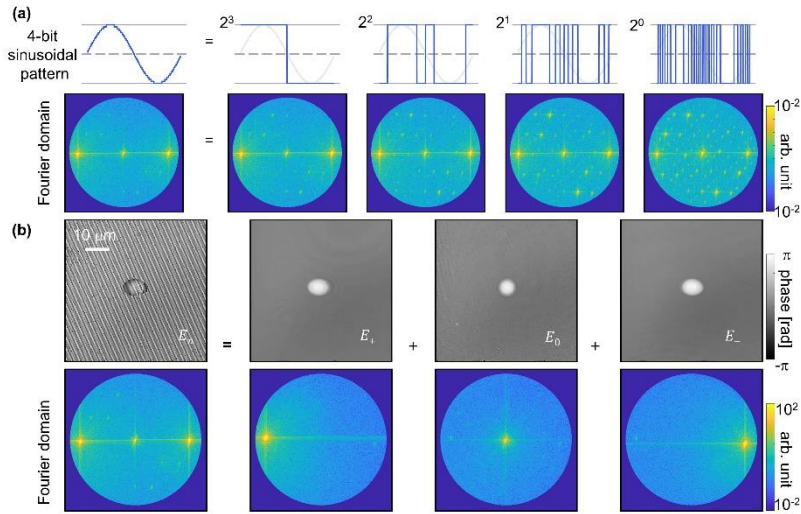

Fig. 2. (a) Digital time-multiplexing scheme for sinusoidal illumination. A single sinusoidal pattern is decomposed into four binary patterns. The collected field information from each binary pattern is numerically compounded. (b) Structured illumination and field decomposition. The scattered field $E_n$ obtained from a sinusoidal illumination pattern is equivalent to the superposition of the scattered fields from the three plane waves $E_+$, $E_-$, and $E_0$.

## 3. Results

### 3.1. Suppression of coherent noise in ODT

To verify the suppression of coherent noise, we compared the measured 3D RI tomograms obtained using the proposed method with the tomograms obtained using conventional ODT with a highly coherent light source. For fair comparison, we used the same optical system, except for the light source. In the system, the SLED and a laser diode (LD) module (CPS450, 450 nm center wavelength, Thorlabs, U.S.) were evaluated as temporally low- and high- coherence sources, respectively. A mirror mounted on a magnetic plate was installed before PBS1 to switch the light source between the SLED and the LD module.

The results are shown in Fig. 3. A polystyrene bead (78462, Sigma-Aldrich, U.S.) with a diameter of 7 μm was immersed in a high RI medium liquid ($n_m$ = 1.56, 1890X, Cargille, USA) and was used as the target sample. When a coherent source was used, parasitic fringes are observable and considerably degrade the image quality [Fig. 3(a) and (d)]. In contrast, the results obtained using the proposed method demonstrate a clear background with no parasitic fringes, and hence, have high image quality. To quantitatively analyze the reduction of coherent noise, the phase delay and RI values in the background area (dotted boxes, Fig. 3) were measured. The standard deviations of the phase values are 113 mrad and 76 mrad for the LD and SLED, respectively. The mean and standard deviations of the RI values are $8\times10^{-4}$ and $6\times10^{-4}$ for the LD and SLED, respectively.

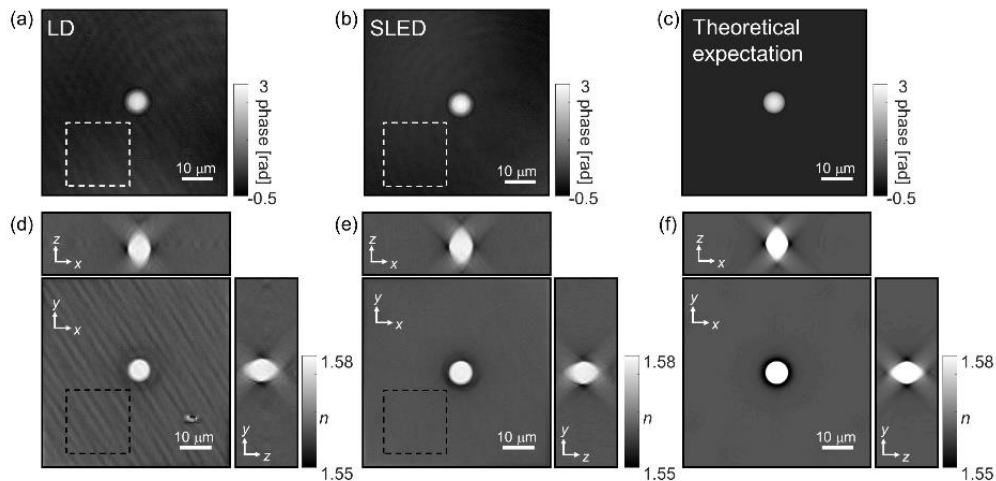

Fig. 3. Coherent noise reduction in ODT. A polystyrene bead with a diameter of 7 μm was measured. The retrieved phase images (from normal illumination) from the (a) LD, (b) SLED, and (c) theoretical expectation. (d)–(f) show the cross-sectional distributions of the reconstructed 3D RI tomograms from the LD, SLED, and theoretical expectation, respectively.

We also calculated the theoretical expectation using the optimal optical transfer function (OTF) of the current system [37] [Figs. 3(c) and 3(f)]. Our results are compatible with the theoretical exception in which no noise was considered. This clearly demonstrates that the proposed method almost completely eliminates the coherent noise in ODT.

### 3.2. Applications to biological samples

The feasibility of the current setup for imaging biological samples was tested by measuring the 3D RI images of various epithelial cells, including Chinese hamster ovary (CHO) [Fig. 4(a)] and NIH 3T3 cells [Fig. 4(b)]. To fill in the missing spectrum in the Fourier domain, the Gerchberg–Saxton algorithm with a non-negativity constraint was used [36] with 25 iterations. Figures 4(a) and 4(b) show the subcellular structures of the epithelial cells without coherent noise at various axial positions. The overall cellular shapes as well as subcellular organelles such as the nucleus membrane and nucleoli are clearly visualized.

To demonstrate the dynamic imaging capability of the proposed method, time-lapsed 3D RI tomograms of live cells were recorded at high speed. K562 cancer cells and CAR T cells were observed with sub-second temporal resolution. The exposure time and the number of fields used for a single 3D RI image were 100 μs and 21, respectively. The 3D RI distributions of the samples were observed for 3.3 s at an acquisition rate of 20.7 Hz (or equivalently, a time interval of 48.4 ms). The dynamics of an individual CAR T cell [Fig. 4(c)] and the subcellular structure of a single K562 cancer cell [red arrows in Fig. 4(d)] were observed at a temporal resolution of tens of milliseconds.

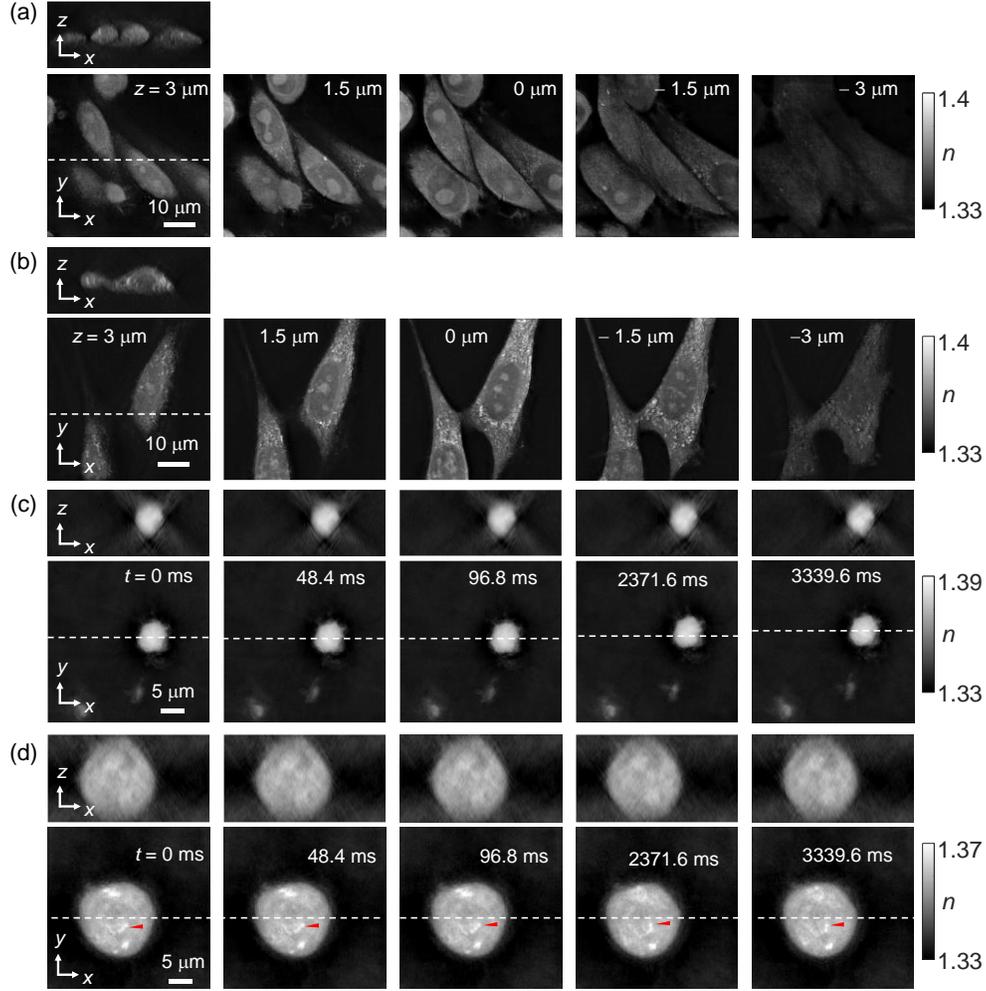

Fig. 4. (a),(b) 3D RI distribution of (a) Chinese hamster ovary (CHO) and (b) NIH 3T3 cells at various axial positions. Intracellular structures are observable with reduced coherent noise. (c),(d) Time-lapse 3D RI distributions of a live (c) CAR T cell and (d) K562 cancer cell.

## 3.3. Spatial resolution

We quantified the spatial resolution by measuring the point spread function (PSF) of the optical setup. The PSF of the system, $P(\mathbf{r})$, was obtained by Wiener deconvolution of the observed scattering potential and the ground truth of a spherical microsphere [38]

$$P(\mathbf{r}) = \mathbf{iFT}\left[\frac{\tilde{S}_0^*(\mathbf{k})}{\left|\tilde{S}_0(\mathbf{k})\right|^2 + \sigma} \tilde{S}_{\mathbf{obs}}(\mathbf{k})\right], \quad (2)$$

where $\tilde{S}_{\mathbf{obs}}(\mathbf{k})$ and $\tilde{S}_0(\mathbf{k})$ are the Fourier spectra of the observed scattering pattern and the ground truth, respectively. The superscript $^*$ and iFT denote complex conjugation and the inverse Fourier transformation, respectively. The value $\sigma$ corresponds to the SNR of the optical system.

As a resolution target, a silica bead with a precise diameter of 4.3 μm (SiO2MS-2.0, Cospheric LLC., U.S.) was immersed in an optical adhesive (NOA 139, Norland Products, USA) and UV-cured. The measured PSF of the setup is shown in Fig.

5(a). The full width at half maximum (FWHM) of the experimental PSF along the *x*-, *y*-, and *z*-axes are 0.23 μm, 0.26 μm, and 0.86 μm, respectively. These results are in good agreement with the theoretical PSF [Fig. 5(b)] calculated using the ideal OTF of the given illumination angle-scanning pattern. The FWHM of the theoretical PSF along the *x*-, *y*-, and *z*-axes are 0.19 μm, 0.19 μm, and 0.59 μm, respectively [Figs. 5(c)–5(d)]. The asymmetry in the PSF is believed to stem from misalignments and aberrations.

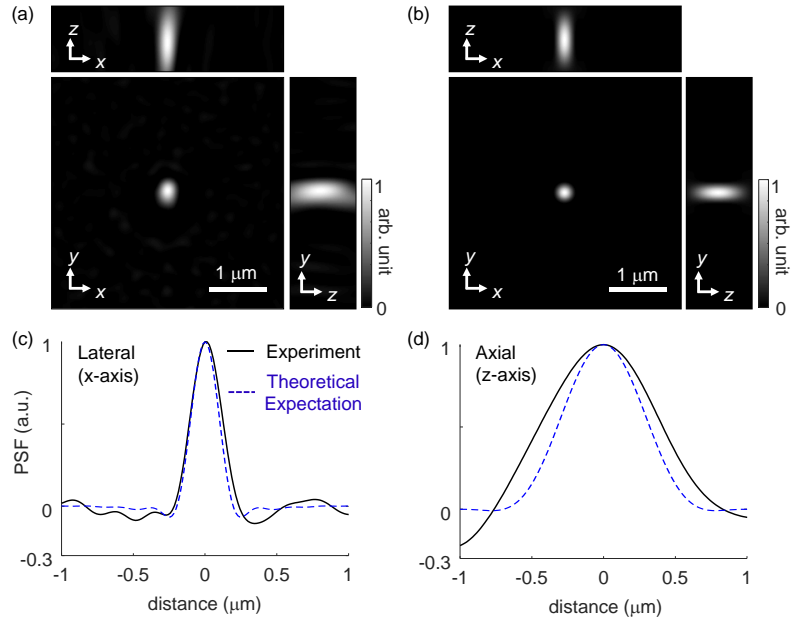

Fig. 5. (a),(b) Cross-sectional image of the (a) experimental and (b) theoretical point spread function (PSF). (c),(d) Line profile of the PSF along the (c) axial (*x*-axis), and (d) lateral (*y*-axis) plane.

## 4. Discussion and Conclusion

We presented a high-speed low-coherence angle-scanning ODT using an FLC SLM. The use of the FLC SLM allowed us to perform fast angle scanning in a diffraction geometry while maintaining the coherence over a large field of view in a simple and compact setup.

We demonstrated the significantly decreased coherent noise in the proposed approach by comparing the 3D RI distributions of a microsphere obtained using SLED and LD modules. Furthermore, we verified the feasibility of the proposed setup by observing the 3D morphology of adherent epithelial cells. Rapid time-lapse imaging of K562 and CAR T cells was also performed to demonstrate the capability for fast 3D imaging. The experimental and theoretical resolutions of the proposed setup were investigated using the Wiener deconvolution method and the optimal OTF of the system, respectively.

We believe that the signal-to-noise ratio and spatiotemporal resolution of the proposed method can be further enhanced by introducing a more powerful light source and optimal device control. For example, in the present demonstration, the maximum frame rate was limited by the illumination power of the SLED, which was far below the speed of the image sensor used. In addition, in principle, only two patterned illuminations are required for the time-multiplexed illumination. However, owing to the uncontrolled residual light from the FLC SLM, we used three patterned illuminations. Ideally, the acquisition speed can be further enhanced by 15 times if the devices are operated at their full capacities, such that a single 3D RI tomogram can be obtained in 80 ms (with 101 complex field images).

The imaging capability of the proposed optical system may be helpful for the exploration of samples with rapid dynamics. For example, fruitful information on protein pathways can be obtained by tracking the rapid movement of primary cilia [39]. The interactions between various intracellular elements, including mitochondria [40], microtubules [41], and cytoskeletons [42] can also be investigated through previously unobtainable views of these interactions.

**Funding**

This work was supported by KAIST, BK21+ program, KAIST Advanced Institute for Science-X, Tomocube, National Research Foundation of Korea (NRF) (2017M3C1A3013923, 2015R1A3A2066550, 2018K000396, 2018R1A6A3A01011043)

**Disclosures**

The authors declare no competing financial interests.

## References

1. Y. Park, C. Depeursinge, and G. Popescu, "Quantitative phase imaging in biomedicine," Nat Photonics **12**, 578-589 (2018).
2. E. Wolf, "Three-dimensional structure determination of semi-transparent objects from holographic data," Optics communications **1**, 153-156 (1969).
3. W. H. Choi, Y. Yun, S. Park, J. H. Jeon, J. Lee, J. H. Lee, S.-A. Yang, N.-K. Kim, C. H. Jung, and Y. T. Kwon, "Aggresomal sequestration and STUB1-mediated ubiquitylation during mammalian proteaphagy of inhibited proteasomes," Proceedings of the National Academy of Sciences (2020).
4. J. Guillén-Boixet, A. Kopach, A. S. Holehouse, S. Wittmann, M. Jahnel, R. Schlüssler, K. Kim, I. R. Trussina, J. Wang, and D. Mateju, "RNA-induced conformational switching and clustering of G3BP drive stress granule assembly by condensation," Cell **181**, 346-361. e317 (2020).
5. M. Baczewska, K. Eder, S. Ketelhut, B. Kemper, and M. Kujawińska, "Holotomographic investigation of an influence of PFA cell fixation process on refractive index of cellular organelles in epithelial cells," in *Quantitative Phase Imaging VI*(International Society for Optics and Photonics2020), p. 112491L.
6. K. Kim, J. Yoon, S. Shin, S. Lee, S.-A. Yang, and Y. Park, "Optical diffraction tomography techniques for the study of cell pathophysiology," Journal of Biomedical Photonics & Engineering **2**, 020201 (2016).
7. S. Kwon, Y. Lee, Y. Jung, J. H. Kim, B. Baek, B. Lim, J. Lee, I. Kim, and J. Lee, "Mitochondria-targeting indolizino[3,2-c]quinolines as novel class of photosensitizers for photodynamic anticancer activity," Eur J Med Chem **148**, 116-127 (2018).
8. J. H. Ahn, H. Seo, W. Park, J. Seok, J. A. Lee, W. J. Kim, G. B. Kim, K.-J. Kim, and S. Y. Lee, "Enhanced succinic acid production by Mannheimia employing optimal malate dehydrogenase," Nature Communications **11**, 1-12 (2020).
9. J. Oh, G.-H. Lee, J. Rho, S. Shin, B. J. Lee, Y. Nam, and Y. Park, "Optical Measurements of Three-Dimensional Microscopic Temperature Distributions Around Gold Nanorods Excited by Localized Surface Plasmon Resonance," Phys Rev Appl **11**, 044079 (2019).
10. F. Charriere, A. Marian, F. Montfort, J. Kuehn, T. Colomb, E. Cuche, P. Marquet, and C. Depeursinge, "Cell refractive index tomography by digital holographic microscopy," Opt Lett **31**, 178-180 (2006).
11. M. Habaza, B. Gilboa, Y. Roichman, and N. T. Shaked, "Tomographic phase microscopy with 180 rotation of live cells in suspension by holographic optical tweezers," Optics letters **40**, 1881-1884 (2015).
12. T. H. Nguyen, M. E. Kandel, M. Rubessa, M. B. Wheeler, and G. Popescu, "Gradient light interference microscopy for 3D imaging of unlabeled specimens," Nature communications **8**, 1-9 (2017).
13. Z. Wang, D. L. Marks, P. S. Carney, L. J. Millet, M. U. Gillette, A. Mihi, P. V. Braun, Z. Shen, S. G. Prasanth, and G. Popescu, "Spatial light interference tomography (SLIT)," Opt Express **19**, 19907-19918 (2011).
14. J. M. Soto, J. A. Rodrigo, and T. Alieva, "Label-free quantitative 3D tomographic imaging for partially coherent light microscopy," Opt Express **25**, 15699-15712 (2017).
15. L. Tian, J. Wang, and L. Waller, "3D differential phase-contrast microscopy with computational illumination using an LED array," Optics letters **39**, 1326-1329 (2014).
16. Y. Sung, W. Choi, C. Fang-Yen, K. Badizadegan, R. R. Dasari, and M. S. Feld, "Optical diffraction tomography for high resolution live cell imaging," Opt Express **17**, 266-277 (2009).
17. A. Kus, W. Krauze, and M. Kujawinska, "Active limited-angle tomographic phase microscope," J Biomed Opt **20**, 111216 (2015).
18. S. Shin, K. Kim, J. Yoon, and Y. Park, "Active illumination using a digital micromirror device for quantitative phase imaging," Opt Lett **40**, 5407-5410 (2015).
19. T. Colomb, F. Montfort, J. Kühn, N. Aspert, E. Cuche, A. Marian, F. Charrière, S. Bourquin, P. Marquet, and C. Depeursinge, "Numerical parametric lens for shifting, magnification, and complete aberration compensation in digital holographic microscopy," JOSA A **23**, 3177-3190 (2006).
20. L. Miccio, D. Alfieri, S. Grilli, P. Ferraro, A. Finizio, L. De Petrocellis, and S. Nicola, "Direct full compensation of the aberrations in quantitative phase microscopy of thin objects by a single digital hologram," Appl Phys Lett **90**, 041104 (2007).
21. I. Choi, K. Lee, and Y. Park, "Compensation of aberration in quantitative phase imaging using lateral shifting and spiral phase integration," Opt Express **25**, 30771-30779 (2017).
22. T. Chang, Y. Jo, G. Choi, D. Ryu, H. Min, and Y. Park, "Calibration-free quantitative phase imaging using data-driven aberration modeling " arXiv (2020).
23. G. Choi, D. Ryu, Y. Jo, Y. S. Kim, W. Park, H.-s. Min, and Y. Park, "Cycle-consistent deep learning approach to coherent noise reduction in optical diffraction tomography," Opt Express **27**, 4927-4943 (2019).
24. S. Shin, K. Kim, K. Lee, S. Lee, and Y. Park, "Effects of spatiotemporal coherence on interferometric microscopy," Opt Express **25**, 8085-8097 (2017).
25. F. Dubois, M.-L. N. Requena, C. Minetti, O. Monnom, and E. Istasse, "Partial spatial coherence effects in digital holographic microscopy with a laser source," Appl Optics **43**, 1131-1139 (2004).
26. Y. Park, W. Choi, Z. Yaqoob, R. Dasari, K. Badizadegan, and M. S. Feld, "Speckle-field digital holographic microscopy," Opt Express **17**, 12285-12292 (2009).
27. M. Rinehart, Y. Zhu, and A. Wax, "Quantitative phase spectroscopy," Biomedical optics express **3**, 958-965 (2012).
28. K. Lee, S. Shin, Z. Yaqoob, P. T. So, and Y. Park, "Low-coherent optical diffraction tomography by angle-scanning illumination," Journal of biophotonics **12**, e201800289 (2019).
29. V. Lauer, "New approach to optical diffraction tomography yielding a vector equation of diffraction tomography and a novel tomographic microscope," Journal of Microscopy **205**, 165-176 (2002).
30. K. Lee, K. Kim, G. Kim, S. Shin, and Y. Park, "Time-multiplexed structured illumination using a DMD for optical diffraction tomography," Optics letters **42**, 999-1002 (2017).
31. A. Kus, "Illumination-related errors in limited-angle optical diffraction tomography," Appl Opt **56**, 9247-9256 (2017).
32. Y. Liu, C. Ma, Y. Shen, J. Shi, and L. V. Wang, "Focusing light inside dynamic scattering media with millisecond digital optical phase conjugation," Optica **4**, 280-288 (2017).
33. K. Lee, K. Kim, G. Kim, S. Shin, and Y. Park, "Time-multiplexed structured illumination using a DMD for optical diffraction tomography," Opt. Lett. **42**, 999-1002 (2017).
34. S. K. Debnath, and Y. Park, "Real-time quantitative phase imaging with a spatial phase-shifting algorithm," Optics letters **36**, 4677-4679 (2011).
35. K. Kim, H. Yoon, M. Diez-Silva, M. Dao, R. R. Dasari, and Y. Park, "High-resolution three-dimensional imaging of red blood cells parasitized by Plasmodium falciparum and in situ hemozoin crystals using optical diffraction tomography," Journal of biomedical optics **19**, 011005 (2013).
36. J. Lim, K. Lee, K. H. Jin, S. Shin, S. Lee, Y. Park, and J. C. Ye, "Comparative study of iterative reconstruction algorithms for missing cone problems in optical diffraction tomography," Opt Express **23**, 16933-16948 (2015).
37. Y. Sung, and R. R. Dasari, "Deterministic regularization of three-dimensional optical diffraction tomography," J Opt Soc Am A Opt Image Sci Vis **28**, 1554-1561 (2011).

38. N. Wiener, *Extrapolation, interpolation, and smoothing of stationary time series: with engineering applications* (MIT press, 1950).
39. A. L. Blitzer, L. Panagis, G. L. Gusella, J. Danias, M. Mlodzik, and C. Iomini, "Primary cilia dynamics instruct tissue patterning and repair of corneal endothelium," Proceedings of the National Academy of Sciences **108**, 2819-2824 (2011).
40. J. Bereiter-Hahn, and M. Vöth, "Dynamics of mitochondria in living cells: shape changes, dislocations, fusion, and fission of mitochondria," Microscopy research and technique **27**, 198-219 (1994).
41. L. Cassimeris, N. K. Pryer, and E. Salmon, "Real-time observations of microtubule dynamic instability in living cells," The Journal of cell biology **107**, 2223-2231 (1988).
42. A. Y. Cheung, Q. H. Duan, S. S. Costa, B. H. de Graaf, V. S. Di Stilio, J. Feijo, and H. M. Wu, "The dynamic pollen tube cytoskeleton: live cell studies using actin-binding and microtubule-binding reporter proteins," Mol Plant **1**, 686-702 (2008).